\documentclass[runningheads]{llncs}
\usepackage{graphicx}
\usepackage{comment}

\usepackage{xcolor}
\PassOptionsToPackage{hyphens,spaces}{url}
\usepackage[colorlinks,urlcolor=blue]{hyperref}
\usepackage{todonotes}
\usepackage{xspace}
\usepackage{bm}
\usepackage[utf8]{inputenc}
\usepackage{amssymb}
\usepackage{amsfonts}
\usepackage{amsmath}

\usepackage{graphicx}
\usepackage{booktabs}
\usepackage{footnoterange}
\usepackage{multirow}

\usepackage[ruled,linesnumbered,vlined]{algorithm2e}

\usepackage{siunitx}

  \DeclareMathOperator{\rev}{rev}

\makeatletter
\renewcommand\section{\@startsection{section}{1}{\z@}%
                       {-12\p@ \@plus -4\p@ \@minus -4\p@}%
                       {8\p@ \@plus 4\p@ \@minus 4\p@}%
                       {\normalfont\large\bfseries\boldmath
                        \rightskip=\z@ \@plus 8em\pretolerance=10000 }}
\makeatother

\begin{document}
\title{Learning Theorem Proving Components
}
\titlerunning{Learning Theorem Proving Components}
\author{Karel Chvalovsk\'y\inst{1}\orcidID{0000-0002-0541-3889}
   \and  Jan Jakub\r{u}v\inst{1,2}\orcidID{0000-0002-8848-5537}
   \and  Miroslav Ol\v{s}\'ak\inst{2}\orcidID{0000-0002-9361-1921}
   \and Josef Urban\inst{1}\orcidID{0000-0002-1384-1613}
}
\authorrunning{K. Chvalovsk\'y et al.}
\titlerunning{Learning Proving Components}
\institute{
   Czech Technical University in Prague, Prague, Czechia \\
   \email{karel@chvalovsky.cz}, \email{josef.urban@gmail.com}
\and
   University of Innsbruck, Innsbruck, Austria\\
   \email{jakubuv@gmail.com}, \email{mirek@olsak.net}
}

\maketitle              %
\begin{abstract}
Saturation-style automated theorem provers (ATPs) based on the
given clause procedure are today the strongest general reasoners for
classical first-order logic.  The clause selection heuristics in such
systems are, however, often evaluating clauses in isolation, ignoring
other clauses.  This has changed recently by equipping the E/ENIGMA
system with a graph neural network (GNN) that chooses the next given
clause based on its evaluation in the context of previously
selected clauses. In this work, we describe several algorithms and
experiments with ENIGMA, advancing the idea of contextual evaluation
based on learning important components of the graph of clauses.
\end{abstract}

\section{Introduction: Clause Selection and Context}
\label{sec:intro}

Clause selection is a crucial part of
saturation-style~\cite{DBLP:books/el/RobinsonV01} automated theorem
provers (ATPs) such as E~\cite{Schulz13}, Vampire~\cite{Vampire}, and
Prover9~\cite{McC-Prover9-URL}.  
These systems, implementing the given-clause~\cite{McCune90} algorithm, provide
the strongest methods for proving lemmas in large interactive theorem
prover (ITP) libraries~\cite{hammers4qed}, and occasionally prove open
conjectures in specialized parts of mathematics~\cite{KinyonVV13}.

Clause selection heuristics have a long history of research, going
back to a number of experiments done with the Otter
system~\cite{McC03-Otter}. Systems such as Prover9 and E have
eventually developed extensive domain-specific languages for clause
selection heuristics, allowing application of sophisticated algorithms based on a
number of different
ideas~\cite{Sch02-AICOMM,Veroff96,Qua92-Book,DBLP:conf/aisc/McCune06,JakubuvU16}
and their automated
improvement~\cite{blistr,SchaferS15,JakubuvU18a}.
These algorithms are, however, often evaluating clauses in isolation,
ignoring other clauses selected in the proof search, and thus largely
neglecting the notion of a \emph{(proof) state} and its obvious
importance for choosing the next \emph{action} (clause).

This has changed recently with equipping the
E/ENIGMA~\cite{JakubuvU17a,ChvalovskyJ0U19} system with a logic-aware graph
neural network (GNN)~\cite{OlsakKU20}, where the next given clause is chosen
based on its evaluation in the context of previously selected
clauses~\cite{JakubuvCOP0U20}.  
In more details, in
GNN-ENIGMA, the generated clauses are not ranked immediately and
independently on other clauses.  Instead, they are judged by the GNN in larger
batches and with respect to a large number of already selected clauses
(\emph{context}). The GNN estimates collectively the most useful
subset of the context and new clauses by several rounds of message passing. The message-passing
algorithm takes into account the connections between symbols, terms, subterms,
atoms, literals, and clauses. It is trained on many previous proof
searches, and it estimates which clauses will collectively benefit the proof
search in the best way.  

In the rest of the paper, we describe several algorithms and
experiments with ENIGMA and GNN-based algorithms, advancing the idea
of contextual evaluation.  In Section~\ref{enigma}, we give an
overview of the learning-based ENIGMA clause selection in E, focusing
on the recently added context-based evaluation by GNNs.
Section~\ref{lfrg} introduces the first variant of our context-based
algorithms called \emph{leapfrogging}. These algorithms interleave
saturation-style ATP runs with external context-based evaluations and
clause filtering.  Section~\ref{components} introduces the second variant
of our context-based algorithms, based on learning from past
interactions between clauses and splitting the proof search into
separate components. Section~\ref{GNN} discusses technical details, and Section~\ref{eval}
evaluates the methods.

\section{ENIGMA and Learning Context-Based Guidance}
\label{enigma}

This section summarizes our previous Graph Neural Network (GNN) ENIGMA
\emph{anonymization} architecture~\cite{JakubuvCOP0U20}, which was previously
successfully used for a given clause guidance within E Prover~\cite{JakubuvU19}.
In this context, anonymization means guidance independent on specific symbol
names.

Saturation-based ATPs, such as E, employ a \emph{given-clause loop}.
The input first-order logic problem is translated into a refutationally-equivalent set of clauses, and a search for a contradiction is initiated.
Starting with the initial set of clauses, one clause is selected (\emph{given})
for processing, and all possible inferences with all previously
processed clauses are derived.
This extends the set of clauses available for processing, and the loop repeats
until (1) the contradiction (empty clause) is derived, or (2) there are no
clauses left for processing (that is, the input problem is not provable), or
(3) resources (time, memory, or user patience) are exhausted.
As the selection of the right clauses for processing is essential for a
success, our approach is to guide the clause selection within an ATP by
sophisticated machine learning methods.

For the clause selection with ENIGMA Anonymous, we train a GNN classifier for
symbol-independent clause embeddings from a large number of previous successful
E proof searches.
From every successful proof search, we extract the set of all processed
clauses, and we label the clauses that appear in the final proof as
\emph{positive} while the remaining (unnecessarily processed) clauses 
as \emph{negative}.
These training data are turned into a tensor representation (one- and
two-dimensional variable-length vectors), which encapsulate clause syntax trees
by abstracting from specific symbol names while preserving information about
symbol relations.
Each tensor represents a set of clauses as a graph with three types of nodes
(for terms/subterms, clauses, and symbols), and passes initial embeddings through a
fixed number of message-passing (graph convolution) layers.
Additionally, the conjecture clauses of the problem to be proved are
incorporated into the graph to allow for conjecture-dependent clause
classification.

Once a GNN classifier is trained from a large number of proof searches, it is
utilized in a new proof search to evaluate the clauses to be processed and to
select the best given clause as follows.
Instead of evaluating the clauses one by one, as is the case in the alternative
ENIGMA Anonymous decision tree classifiers, we postpone clause evaluation until
a specific number of clauses to be evaluated is collected.
These clauses form the \emph{query} part and the size of the query is passed
to the prover as a parameter.
The query clauses are extended with clauses forming a \emph{context}, that is,
a specific number of clauses already processed during the current proof
search.
In particular, we use the first $n$ clauses processed during the proof search as the context.
The context size $n$ is another parameter passed to the prover.
After adding the conjecture and context clauses to the query, their tensor
representation is computed and sent to the GNN for evaluation.
The GNN applies several graph convolution (message passing) layers
getting an embedding of every clause. Each clause is combined
through a single fully connected layer with an embedding of the
conjecture, and finally transformed into a single score (\emph{logit}),
which is sent back to the prover.
The prover then processes the clauses with better
(higher) scores in advance. For details, see~\cite{JakubuvCOP0U20,OlsakKU20}.

\section{Leapfrogging}
\label{lfrg}
The first class of
algorithms is based on the idea that the graph-based evaluation of a
particular clause may significantly change as new clauses are produced
and the context changes. It corresponds to the human-based
mathematical exploration, in which initial actions can be done with
relatively low confidence and following only uncertain hunches. After
some amount of initial exploration is done, clearer patterns often appear,
allowing re-evaluation of the approach, focusing on the most promising
directions, and discarding of less useful ideas.

In tableau-based
provers such as leanCoP~\cite{OB03} with a compact notion of state,
such methods can be approximated in a reinforcement learning setting
by the notion of \emph{big steps}~\cite{KaliszykUMO18} in the
Monte-Carlo tree search (MCTS), implementing the standard
\emph{explore/exploit} paradigm~\cite{gittins1979bandit}.  In the saturation setting, our proposed
algorithm uses short standard saturation runs at the exploration
phase, after which the set of processed (selected) clauses is
reevaluated and a decision on its most useful subset is made by the
GNN. These two phases are iterated in a procedure that we call
\emph{leapfrogging}.

In more detail, leapfrogging is implemented as follows (see also Algorithm~\ref{alg:lfrg}). Given a
clausal problem consisting of a set of initial clauses $S=S_0$, an
initial saturation-style search (in our case E/ENIGMA) is run on $S$
with an abstract time limit. We may use a fixed limit (e.g., 1000
nontrivial processed clauses) for all runs, or change (e.g. increase) the limits gradually. If the initial run results in a proof
or saturation within the limit, the algorithm is finished. If not, we
inspect the set of clauses created in the run. We can inspect the set
of all generated clauses, or a smaller set, such as the set of all
processed clauses. So far, we used the latter because it is typically
much smaller and better suits our training methods. This
(\emph{large}) set is denoted as $L_0$. Then we apply a trained
graph-based predictor to $L_0$, which selects a smaller \emph{most
promising} subset of $L_0$, denoted as $S_1$. We may or may not automatically include also 
the initial negated conjecture clauses or the whole initial set 
$S_0$ in $S_1$. $S_1$ is then used as an input to the next limited saturation run of E/ENIGMA.
This process is iterated, producing gradually sets $S_i$ and $L_i$. 

\setlength{\textfloatsep}{.5em}%
\begin{algorithm}[t]%
  \caption{The Leapfrogging algorithm with a fixed saturation limit}\label{alg:lfrg}
\KwIn{$AxiomClauses$, $NegConjectureClauses$, $SaturationLimit$, $IterationLimit$, $PremiseSelector$\;}
$S_0 = AxiomClauses \cup NegConjectureClauses$\;
\For{$i = 0$ \KwTo $IterationLimit$}{
$(L_{i+1}, Result)$ =      $Saturate(S_i,SaturationLimit)$\;
\lIf {$Result = Unsatisfiable$} {\Return $Unsatisfiable$}
\uElseIf {$Result = Satisfiable$} { \lIf {i=0} {\Return Satisfiable} 
 \lElse { \Return $Unknown$}}
\Else(\tcp*[h]{$Result = Unknown$})
{
  $S_{i+1}$ = $PremiseSelector(L_{i+1},NegConjectureClauses)$ \;
$S_{i+1} = S_{i+1} \cup NegConjectureClauses$\;
}
}
\Return $Unknown$\;
  \end{algorithm}

A particularly simple version of leapfrogging uses GNN-guided ENIGMA
for the saturation ``jumps'', and omits the external selection, thus
setting $S_{i+1}:=L_{i}$. This may seem meaningless with deterministic
clause selection heuristics that do not use context: the next
saturation run may be selecting the same clauses and ending up with
$S_{i+1}=S_i$.  Already in the standard ATP setting this is, however,
easy to make less deterministic, as done, for example, in the randoCoP
system~\cite{RathsO08}. The GNN-guided ENIGMA will typically also make
different choices with the new input set $L_0$ than with the input
set $S_0$.

A more involved version of leapfrogging, however, makes use of a
nontrivial trained graph-based predictor that will reduce $L_i$ to
$S_{i+1}$ such that $S_{i+1} \subsetneq L_i$.  For this, we use an
external evaluation run of a GNN, which has been trained in the same
way as the GNN used inside ENIGMA: on sets of \emph{positive} and
\emph{negative} processed clauses extracted from many successful proof
runs. Here, the positive clauses are those that end up being part of
the proof, and the negative ones are the remaining processed
clauses. This is also very similar to an external \emph{premise
  selection}~\cite{abs-1108-3446} done with the GNNs~\cite{OlsakKU20},
with the difference that the inputs are now clauses instead of
formulas.

\section{Learning Reasoning Components}
\label{components}
The second class of algorithms is based on learning important
components in the graph of clauses. This is again motivated by an
analogy with solving mathematical problems, which often have
well-separated reasoning and computational components. Examples
include numerical calculations, computing derivatives and
integrals, performing Boolean algebra in various settings, sequences
of standard rewriting and normalization operations in various
algebraic theories, etc. Such components of the larger problem can be
often solved mostly in isolation from the other components, and only
their results are then used together to connect them and solve the
larger problem.

Human-designed problem solving architectures addressing such
decomposition include, e.g., SMT systems, systems such as
MetiTarski~\cite{AkbarpourP10}, and a tactic-based learning-guided proof
search in systems such as TacticToe~\cite{GauthierKU17}. In all these
systems, the component procedures or tactics are, however,
\emph{human-designed} and (often painstakingly) human-implemented,
with a lot of care both for the components and for the algorithms that
merge their results. This approach seems hard to scale to the large
number of combinations of complex algorithms, decision procedures and
reasoning heuristics used in research-level mathematics, and other
complex reasoning domains.

Our new approach is to instead start to \emph{learn} such
\emph{targeted components}, expressed as sets of clauses that perform
targeted reasoning and computation within the saturation framework. We
also want to learn the merging of the results of the components
automatically.  This is quite ambitious, but there seems to be 
growing evidence that such targeted components are being learned in
many iterations of GNN-guided proving followed by retraining of the
GNNs in our recent large iterative evaluation over
Mizar.\footnote{\url{https://github.com/ai4reason/ATP_Proofs}} In
these experiments, we have significantly extended our previously
published results~\cite{JakubuvCOP0U20},\footnote{The publication of
  this large evaluation is in preparation.}  eventually automatically
proving 73.5\% (more than \num{40}k) of the Mizar theorems.  In
particular, there are many examples shown on the project Github page
demonstrating that the GNN is learning to solve more and more involved
computations in problems involving differentiation, integration,
boolean algebra, algebraic rewriting, etc. Our initial approach 
is therefore to (i) use the GNN to learn to identify interacting
reasoning components, (ii) use graph-based and clustering-based
algorithms to split the set of clauses into components based on the
GNN predictions, (iii) run saturation on the components independently,
(iv) possibly merge the most important parts of the components, and
(v) iterate. See the Split and Merge Algorithm~\ref{alg:splitmerge}. 

\begin{algorithm}[hptb]
  \caption{The Split and Merge algorithm}\label{alg:splitmerge}
\KwIn{$AxiomClauses$, $NegConjectureClauses$, $SaturationLimit$, $IterationLimit$, $PremiseSelector$, $ClusteringAlgo$\;}
$S_0 = AxiomClauses \cup NegConjectureClauses$\;
\For{$i = 0$ \KwTo $IterationLimit$}{
$(L_i, Result)$ =      $Saturate(S_i,SaturationLimit)$\;
\lIf {$Result = Unsatisfiable$} {\Return $Unsatisfiable$}
\uElseIf {$Result = Satisfiable$} { \lIf {i=0} {\Return Satisfiable} 
 \lElse { \Return $Unknown$}}
\Else(\tcp*[h]{$Result = Unknown$})
{
  $(C^1_i, ..., C^K_i)$ = $ClusteringAlgo(L_i)$ \tcp*{Split to components}  
  \For{$j = 1$ \KwTo $K$}{
    $(L^j_i, Result^j) = Saturate(C^j_i,SaturationLimit)$ \tcp*{Run each}
    \lIf {$Result^j = Unsatisfiable$} {\Return $Unsatisfiable$}
  }
  $S_{i+1}$ = $PremiseSelector(\bigcup\limits_{j=1}^K L^j_i  ,NegConjectureClauses)$ \tcp*{Merge}  
$S_{i+1} = S_{i+1} \cup NegConjectureClauses$\;
}
}
\Return $Unknown$\;
\end{algorithm}

\section{Clustering Methods}
\label{GNN}

Here we propose two modifications of our previous GNN architecture, described
in Section~\ref{enigma}, for the identification of interacting reasoning components,
and we describe their intended use.
The overall methodology to detect and utilize reasoning components is as follows.
To produce the training data, we run E with a fixed limit of $N$ given clause loops.
For each solved problem, we output not only the proof, but the full derivation
tree of all clauses \emph{generated} during the proof search.
These will provide training data to train a GNN classifier.
For unproved problems, we output the $N$ given clauses \emph{processed} during
the search.
These data from unsuccessful proof searches are then used for the prediction of
interacting components. This is the start of the Split and Merge Algorithm~\ref{alg:splitmerge}.

The training data are extracted from successful proof searches as follows.
From each derivation tree, we extract all 
clause pairs $C_i$ and $C_j$ which interacted during the proof search, that is,
the pairs which were used to infer another clause.
All pairs $(C_i,C_j)$ which were used to infer a proof clause are marked as
\emph{positive} while the remaining clause pairs as \emph{negative}.
Such clauses with the information about their positive/negative pairing
are used to train a GNN predictor.

The trained GNN predictor will guide the construction of clusters, where 
clauses resembling positively linked clauses should end up within the same
cluster.  
We obtain the data for predictions from the above unsuccessful proof searches
(with the fixed limit of $N$ processed clauses), and they contain $N$ processed
clauses for every problem.  
We want to assign each pair of clauses $(C_i,C_j)$ a score $l_{i,j}$ which
describes the likelihood of inferring a useful clause from $C_i$
and $C_j$.
These scores are the basis for the clustering algorithms.

We experiment with two slightly different GNN architectures for the
identification of reasoning components.  Let $d$ be the dimension of
the final clause embedding, and let $c_i$, $c_j$ be the embeddings of
clauses $C_i$, $C_j$ respectively.  Then the two architectures ---
differently computing the score $l_{i,j}$ ---  are as follows:
\begin{enumerate}
\item We pass both $c_i, c_j$ through a linear layer (with biases,
  without an activation function) with the output dimension $n$,
  resulting in $d_i, d_j$. Then we calculate
  $l_{i,j} = d_i \cdot d_j / \sqrt n$.
\item
  We pass both $c_i, c_j$ through a linear layer with the
  output dimension $2n$, resulting in $d'_i, d'_j$. Then we calculate
  $l_{i,j} = d_i \cdot \rev(d_j) / \sqrt n$ where $\rev$ represents
  reversing the vector.
\end{enumerate}
Mathematically, this corresponds to
$l_{i,j} = c_i^T A c_j + v^T (c_i+c_j) + b$
where $v^T$ are $n$-dimensional vectors for clause evaluation,
$b$ is a scalar bias, and A is an $n\times n$ matrix which is
symmetric and positive definite in architecture~1, and just symmetric
in architecture~2. For training, we pass the value $l_{i,j}$ through
sigmoid and binary cross entropy loss.

\subsection{Clustering}
\label{sec:clustering}

To split the clauses into separate components, we use
standard clustering algorithms. However, in our case, it is likely
that some clauses should be shared among various components, and hence
we are also interested in methods capable of such overlapping
assignments.

Of course, the crucial precondition for splitting clauses into
components is defining the similarity between clauses, or even better, a
distance between them. We have at least two straightforward options
here---to define the distance between two clauses as the distance
between their embeddings (vectors) or use the matrix $L = (l_{ij})$ as a similarity measure, which approximates the likelihood
that clauses $i$ and $j$ interact in the proof. A simple way to produce
distances from $L$ is to treat each row of $L$ as a vector and define
the distance between two clauses as the (Euclidean) distance between the
corresponding rows of $L$.

Another approach is to use directly the intended meaning of matrix $L$, the
likelihood that two given clauses appear in a proof, and to
produce a weighted graph from $L$, where vertices are clauses and
edges are assigned weights according to $L$. Moreover, we can remove
edges that have weights below some threshold, expressing that such
clauses do not interact. In this way, we obtain a weighted graph that
can be clustered into components.
The following paragraphs briefly describe the clustering algorithms
used in the experiments.

\vspace{-.5em}
\subsubsection{$k$-means}
\label{sec:kmeans}

A widely used clustering method is $k$-means. The goal is to separate
vectors into $k$ clusters in such a way that their within-cluster
variance is minimal. Although $k$-means is a popular clustering
method, it suffers from numerous well-known problems. For example, it
assumes that we know the correct number of clusters in advance, the
clusters are of similar sizes, and they are nonoverlapping. Although
these assumptions are not satisfied in our case, we used
$k$-means from SciPY~\cite{2020SciPy-NMeth} as a well-known baseline.

\vspace{-.5em}
\subsubsection{Soft $k$-means}
\label{sec:soft-kmeans}

It is possible to modify $k$-means in such a way that overlapping
clusters (also called soft clusters) are allowed.\footnote{Another
  popular way how to generalize $k$-means (and assign a point to more
  than one cluster) is to use Gaussian mixture models.} An example is
the Fuzzy C-Means (FCM) algorithm~\cite{Bezdek1984} that generalizes
$k$-means by adding the membership function for each point. This
function scores how much each point belongs to a cluster, and it is
possible to adjust the degree of overlap between the clusters. We used the
\texttt{fuzzy-c-means} package~\cite{Dias2019fuzzy} for our experiments.

\vspace{-.5em}
\subsubsection{Graph clustering}
\label{sec:graph-clustering}

We have experimented with the cluster application from the popular
Graphviz visualisation software~\cite{DBLP:books/sp/04/EllsonGKNW04},
which can split a graph into clusters using the methods described
in~\cite{Blondel2008fast}. The graphs are clustered based on the
\emph{modularity} measure which considers the density of links inside a cluster
compared to links between clusters.  It is possible to either directly
specify the intended number of clusters (soft constraint), or base the
number of clusters on their modularity.  We also experimented with
clustering using the \emph{modularity quality}.\footnote{\url{https://gitlab.com/graphviz/graphviz/-/blob/main/lib/sparse/mq.h}}
Moreover, by removing some highly connected vertices (clauses)
before clustering and adding them into all clusters, we can produce
overlapping clusters.

\section{Evaluation}
\label{eval}

\subsection{Leapfrogging}
The first leapfrogging experiment is done as follows:
\begin{enumerate}
\item We stop GNN-ENIGMA after 300 processed clauses and print them.
\item We restart with the 300 clauses used as input, stop at 500 clauses and print the 500 clauses.
\item We restart with the 500 clauses, and do a final run for 60s.
\end{enumerate}

This is done on a set of 28k \emph{hard} Mizar problems that we have
been trying to prove with many different methods in a large ongoing
evaluation over the full Mizar corpus.\footnote{Details are at
  \url{https://github.com/ai4reason/ATP_Proofs}}
We try with four differently trained and parameterized GNNs, denoted as $G_1$, $\dots$, $G_4$.\footnote{The detailed parameters of the GNNs are given in Appendix~\ref{app:data}.} The summary of the runs is given in Table~\ref{lfrg1}.

\begin{table}[t]
\begin{small}
  \caption{\label{lfrg1} Four leapfrogging runs with different GNN-ENIGMAs}
  \centering
    \vspace*{-1em}
  \begin{tabular}{lllll}
    \toprule
    GNN-strategy & original-60s-run &  leapfrogging (300-500-60s) & union & added-by-lfrg \\
    \midrule
    $G_1$ & 2711 & 2218 & 3370 & 659 \\
    $G_2$ & 2516 & 2426 & 3393 & 877 \\
    $G_3$ & 2655 & 2463 & 3512 & 857 \\
             $G_4$ & 2477 & 2268 & 3276 & 799 \\
       \bottomrule
  \end{tabular}
  \end{small}
\end{table}
We see that the methods indeed achieve high complementarity to the
original GNN strategies. This is most likely thanks to the different context
in which the GNN sees the initial clauses in the subsequent runs.

\subsection{Splitting and Merging }
The initial experimental evaluation\footnote{On a server with 36 hyperthreading Intel(R)
Xeon(R) Gold 6140 CPU @ 2.30GHz cores, 755 GB of memory, and 4 NVIDIA GeForce
GTX 1080 Ti GPUs.}
is done on a large benchmark of $57880$ Mizar40~\cite{KaliszykU13b} problems%
\footnote{\url{http://grid01.ciirc.cvut.cz/~mptp/1147/MPTP2/problems_small_consist.tar.gz}}
exported to first-order logic by MPTP~\cite{Urban06}.
We use a subset of 52k Mizar40~\cite{KaliszykU13b} problems for training.
To produce the training data, we run E with a well-performing GNN
guidance, and with the limit of \num{1000} given clause loops.
Within this limit, around \num{20}k of the training problems are solved.
For the \num{32}k unproved training problems, we output the \num{1000} given clauses
\emph{processed} during the search.
As described in Section~\ref{GNN}, we train a GNN predictor on the \num{20}k successful
runs, and use it  to predict the interactions between the processed clauses of the unsuccessful runs.
\begin{table}[t]
\begin{center}
\begin{small}
  \caption{Clustering 3000 problems for evaluation}\label{tclustering}
  \centering
  \vspace*{-1em}
  \begin{tabular}{lcc}
    \toprule
    Method & \#clusters & Newly solved problems\\
    \midrule
    $k$-means & 2 & 67 \\
    $k$-means & 3 & 78 \\
    soft $k$-means & 2 & 63\\
    soft $k$-means & 3 & 93 \\
    Graphviz & $\leq 4$ & 111\\
    \bottomrule
  \end{tabular}
  \vspace*{-1em}
  \end{small}
\end{center}
\end{table}
Since the evaluation on the full set of the \num{32}k unsolved
problems would be too resource-intensive, we limit this to its randomly
chosen \num{3000}-big subset. Table~\ref{tclustering}
shows the performance of the clustering methods in solving the problems in the
first Split phase. The strongest method is the Graphviz-based graph clustering.
In more detail,  %
the \texttt{cluster}
tool gives us on the GNN graph predictions up to four graph
components. 
We run again with a 1000-given clause limit on them newly solving altogether
\num{111} problems inside the components of the \num{3000}.

Then we choose this clustering for an experiment with the Merge phase. 
We merge the
components of the remaining unsolved 2889 problems and use our GNN for a premise-selection-style final choice
of the jointly best subset of the clauses produced by all the
components (line 13 of Algorithm~\ref{alg:splitmerge}). We use four thresholds for the premise selection, and run
again with a 1000-given clause limit on each of such premise
selections %
(line 3 of Algorithm~\ref{alg:splitmerge}).
This run on the merged components yields another 66 new proofs.
Many of the newly found proofs indeed show frequent computational patterns. Examples (see also Appendix~\ref{app:data}) include
the proofs of Mizar problems
\texttt{T16\_FDIFF\_5},\footnote{\url{http://grid01.ciirc.cvut.cz/~mptp/7.13.01_4.181.1147/html/fdiff_5.html\#T16}}
\texttt{T48\_NEWTON},\footnote{\url{http://grid01.ciirc.cvut.cz/~mptp/7.13.01_4.181.1147/html/newton.html\#T48}}
\texttt{T10\_MATRIX\_4},\footnote{\url{http://grid01.ciirc.cvut.cz/~mptp/7.13.01_4.181.1147/html/matrix_4.html\#T10}}
\texttt{T11\_VECTSP\_2},\footnote{\url{http://grid01.ciirc.cvut.cz/~mptp/7.13.01_4.181.1147/html/vectsp_2.html\#T11}}
\texttt{T125\_RVSUM\_1},\footnote{\url{http://grid01.ciirc.cvut.cz/~mptp/7.13.01_4.181.1147/html/rvsum_1.html\#T125}}
\texttt{T13\_BCIALG\_3},\footnote{\url{http://grid01.ciirc.cvut.cz/~mptp/7.13.01_4.181.1147/html/bcialg_3.html\#T13}}
and \texttt{T14\_FUZZY\_2}.\footnote{\url{http://grid01.ciirc.cvut.cz/~mptp/7.13.01_4.181.1147/html/fuzzy_2.html\#T14}}

\section{Conclusion}
We have described several algorithms advancing the idea of contextual
evaluation based on learning important components of the graph of
clauses. The first leapfrogging experiments already show very
encouraging results on the Mizar dataset, providing many complementary
solutions. The component-based algorithm also produces new solutions
and there are clearly many further methods and experiments that can be
tried in this setting.  We believe that this approach may eventually
lead to using large mathematical libraries for automated learning of
nontrivial components, algorithms, and decision procedures involved in
mathematical reasoning.

\section*{Acknowledgments}

Supported by the ERC Consolidator grant \emph{AI4REASON} no.~649043 %
(JJ, JU), by the Czech project AI \& Reasoning
CZ.02.1.01/0.0/0.0/15\_003/0000466 and the European Regional Development Fund
(KC, JU), by the ERC Starting grant \emph{SMART} no.~714034 (JJ, MO),
and by the Czech MEYS under the ERC CZ project \emph{POSTMAN} no. LL1902 (JJ).

\bibliographystyle{plain}

\bibliography{main}

\begin{thebibliography}{10}

\bibitem{AkbarpourP10}
Behzad Akbarpour and Lawrence~C. Paulson.
\newblock {MetiTarski}: An automatic theorem prover for real-valued special
  functions.
\newblock {\em J. Autom. Reasoning}, 44(3):175--205, 2010.

\bibitem{abs-1108-3446}
Jesse Alama, Tom Heskes, Daniel K\"{u}hlwein, Evgeni Tsivtsivadze, and Josef
  Urban.
\newblock Premise selection for mathematics by corpus analysis and kernel
  methods.
\newblock {\em J. Autom. Reasoning}, 52(2):191--213, 2014.

\bibitem{Bezdek1984}
James~C. Bezdek, Robert Ehrlich, and William Full.
\newblock Fcm: The fuzzy c-means clustering algorithm.
\newblock {\em Computers \& Geosciences}, 10(2):191--203, 1984.

\bibitem{hammers4qed}
Jasmin~Christian Blanchette, Cezary Kaliszyk, Lawrence~C. Paulson, and Josef
  Urban.
\newblock Hammering towards {QED}.
\newblock {\em J. Formalized Reasoning}, 9(1):101--148, 2016.

\bibitem{Blondel2008fast}
Vincent~D Blondel, Jean-Loup Guillaume, Renaud Lambiotte, and Etienne Lefebvre.
\newblock Fast unfolding of communities in large networks.
\newblock {\em Journal of statistical mechanics: theory and experiment},
  2008(10):P10008, 2008.

\bibitem{ChvalovskyJ0U19}
Karel Chvalovsk{\'{y}}, Jan Jakub\r{u}v, Martin Suda, and Josef Urban.
\newblock {ENIGMA-NG:} efficient neural and gradient-boosted inference guidance
  for {E}.
\newblock In Pascal Fontaine, editor, {\em Automated Deduction - {CADE} 27 -
  27th International Conference on Automated Deduction, Natal, Brazil, August
  27-30, 2019, Proceedings}, volume 11716 of {\em Lecture Notes in Computer
  Science}, pages 197--215. Springer, 2019.

\bibitem{Dias2019fuzzy}
Madson Luiz~Dantas Dias.
\newblock fuzzy-c-means: An implementation of fuzzy $c$-means clustering
  algorithm., 2019.

\bibitem{DBLP:books/sp/04/EllsonGKNW04}
John Ellson, Emden~R. Gansner, Eleftherios Koutsofios, Stephen~C. North, and
  Gordon Woodhull.
\newblock Graphviz and dynagraph - static and dynamic graph drawing tools.
\newblock In Michael J{\"{u}}nger and Petra Mutzel, editors, {\em Graph Drawing
  Software}, pages 127--148. Springer, 2004.

\bibitem{GauthierKU17}
Thibault Gauthier, Cezary Kaliszyk, and Josef Urban.
\newblock {TacticToe}: Learning to reason with {HOL4} tactics.
\newblock In Thomas Eiter and David Sands, editors, {\em LPAR-21, 21st
  International Conference on Logic for Programming, Artificial Intelligence
  and Reasoning, Maun, Botswana, May 7-12, 2017}, volume~46 of {\em EPiC Series
  in Computing}, pages 125--143. EasyChair, 2017.

\bibitem{gittins1979bandit}
John~C Gittins.
\newblock Bandit processes and dynamic allocation indices.
\newblock {\em J. the Royal Statistical Society. Series B (Methodological)},
  pages 148--177, 1979.

\bibitem{DBLP:conf/gcai/2015}
Georg Gottlob, Geoff Sutcliffe, and Andrei Voronkov, editors.
\newblock {\em Global Conference on Artificial Intelligence, {GCAI} 2015,
  Tbilisi, Georgia, October 16-19, 2015}, volume~36 of {\em EPiC Series in
  Computing}. EasyChair, 2015.

\bibitem{JakubuvCOP0U20}
Jan Jakub\r{u}v, Karel Chvalovsk{\'{y}}, Miroslav Ol\v{s}{\'{a}}k, Bartosz
  Piotrowski, Martin Suda, and Josef Urban.
\newblock {ENIGMA} anonymous: Symbol-independent inference guiding machine
  (system description).
\newblock In Nicolas Peltier and Viorica Sofronie{-}Stokkermans, editors, {\em
  Automated Reasoning - 10th International Joint Conference, {IJCAR} 2020,
  Paris, France, July 1-4, 2020, Proceedings, Part {II}}, volume 12167 of {\em
  Lecture Notes in Computer Science}, pages 448--463. Springer, 2020.

\bibitem{JakubuvU16}
Jan Jakub\r{u}v and Josef Urban.
\newblock Extending {E} prover with similarity based clause selection
  strategies.
\newblock In Michael Kohlhase, Moa Johansson, Bruce~R. Miller, Leonardo
  de~Moura, and Frank~Wm. Tompa, editors, {\em Intelligent Computer Mathematics
  - 9th International Conference, {CICM} 2016, Bialystok, Poland, July 25-29,
  2016, Proceedings}, volume 9791 of {\em Lecture Notes in Computer Science},
  pages 151--156. Springer, 2016.

\bibitem{JakubuvU17a}
Jan Jakub\r{u}v and Josef Urban.
\newblock {ENIGMA:} efficient learning-based inference guiding machine.
\newblock In Herman Geuvers, Matthew England, Osman Hasan, Florian Rabe, and
  Olaf Teschke, editors, {\em Intelligent Computer Mathematics - 10th
  International Conference, {CICM} 2017, Edinburgh, UK, July 17-21, 2017,
  Proceedings}, volume 10383 of {\em Lecture Notes in Computer Science}, pages
  292--302. Springer, 2017.

\bibitem{JakubuvU18a}
Jan Jakub\r{u}v and Josef Urban.
\newblock Hierarchical invention of theorem proving strategies.
\newblock {\em {AI} Commun.}, 31(3):237--250, 2018.

\bibitem{JakubuvU19}
Jan Jakub\r{u}v and Josef Urban.
\newblock Hammering {Mizar} by learning clause guidance.
\newblock In John Harrison, John O'Leary, and Andrew Tolmach, editors, {\em
  10th International Conference on Interactive Theorem Proving, {ITP} 2019,
  September 9-12, 2019, Portland, OR, {USA}}, volume 141 of {\em LIPIcs}, pages
  34:1--34:8. Schloss Dagstuhl - Leibniz-Zentrum f{\"{u}}r Informatik, 2019.

\bibitem{KaliszykU13b}
Cezary Kaliszyk and Josef Urban.
\newblock {MizAR 40 for Mizar 40}.
\newblock {\em J. Autom. Reasoning}, 55(3):245--256, 2015.

\bibitem{KaliszykUMO18}
Cezary Kaliszyk, Josef Urban, Henryk Michalewski, and Miroslav~Ol\v s{\'{a}}k.
\newblock Reinforcement learning of theorem proving.
\newblock In {\em Advances in Neural Information Processing Systems 31: Annual
  Conference on Neural Information Processing Systems 2018, NeurIPS 2018, 3-8
  December 2018, Montr{\'{e}}al, Canada.}, pages 8836--8847, 2018.

\bibitem{KinyonVV13}
Michael~K. Kinyon, Robert Veroff, and Petr Vojtechovsk{\'{y}}.
\newblock Loops with abelian inner mapping groups: An application of automated
  deduction.
\newblock In Maria~Paola Bonacina and Mark~E. Stickel, editors, {\em Automated
  Reasoning and Mathematics - Essays in Memory of William W. McCune}, volume
  7788 of {\em LNCS}, pages 151--164. Springer, 2013.

\bibitem{Vampire}
Laura Kov{\'a}cs and Andrei Voronkov.
\newblock First-order theorem proving and {V}ampire.
\newblock In Natasha Sharygina and Helmut Veith, editors, {\em CAV}, volume
  8044 of {\em LNCS}, pages 1--35. Springer, 2013.

\bibitem{McCune90}
William McCune.
\newblock {OTTER} 2.0.
\newblock In Mark~E. Stickel, editor, {\em 10th International Conference on
  Automated Deduction, Kaiserslautern, FRG, July 24-27, 1990, Proceedings},
  volume 449 of {\em Lecture Notes in Computer Science}, pages 663--664.
  Springer, 1990.

\bibitem{McC-Prover9-URL}
William McCune.
\newblock {Prover9 and Mace4}.
\newblock \url{http://www.cs.unm.edu/~mccune/prover9/}, 2005--2010.

\bibitem{DBLP:conf/aisc/McCune06}
William McCune.
\newblock Semantic guidance for saturation provers.
\newblock In Jacques Calmet, Tetsuo Ida, and Dongming Wang, editors, {\em
  Artificial Intelligence and Symbolic Computation, 8th International
  Conference, {AISC} 2006, Beijing, China, September 20-22, 2006, Proceedings},
  volume 4120 of {\em Lecture Notes in Computer Science}, pages 18--24.
  Springer, 2006.

\bibitem{McC03-Otter}
W.W. McCune.
\newblock {Otter 3.3 Reference Manual}.
\newblock Technical Report ANL/MSC-TM-263, Argonne National Laboratory,
  Argonne, USA, 2003.

\bibitem{OlsakKU20}
Miroslav Ol\v{s}{\'{a}}k, Cezary Kaliszyk, and Josef Urban.
\newblock Property invariant embedding for automated reasoning.
\newblock In Giuseppe~De Giacomo, Alejandro Catal{\'{a}}, Bistra Dilkina,
  Michela Milano, Sen{\'{e}}n Barro, Alberto Bugar{\'{\i}}n, and
  J{\'{e}}r{\^{o}}me Lang, editors, {\em {ECAI} 2020 - 24th European Conference
  on Artificial Intelligence, 29 August-8 September 2020, Santiago de
  Compostela, Spain, August 29 - September 8, 2020 - Including 10th Conference
  on Prestigious Applications of Artificial Intelligence {(PAIS} 2020)}, volume
  325 of {\em Frontiers in Artificial Intelligence and Applications}, pages
  1395--1402. {IOS} Press, 2020.

\bibitem{OB03}
Jens Otten and Wolfgang Bibel.
\newblock {leanCoP:} lean connection-based theorem proving.
\newblock {\em J. Symb. Comput.}, 36(1-2):139--161, 2003.

\bibitem{Qua92-Book}
Art Quaife.
\newblock {\em {Automated Development of Fundamental Mathematical Theories}}.
\newblock Kluwer Academic Publishers, 1992.

\bibitem{RathsO08}
Thomas Raths and Jens Otten.
\newblock randocop: Randomizing the proof search order in the connection
  calculus.
\newblock In Boris Konev, Renate~A. Schmidt, and Stephan Schulz, editors, {\em
  Proceedings of the First International Workshop on Practical Aspects of
  Automated Reasoning, Sydney, Australia, August 10-11, 2008}, volume 373 of
  {\em {CEUR} Workshop Proceedings}. CEUR-WS.org, 2008.

\bibitem{DBLP:books/el/RobinsonV01}
John~Alan Robinson and Andrei Voronkov, editors.
\newblock {\em Handbook of Automated Reasoning (in 2 volumes)}.
\newblock Elsevier and {MIT} Press, 2001.

\bibitem{SchaferS15}
Simon Sch{\"{a}}fer and Stephan Schulz.
\newblock Breeding theorem proving heuristics with genetic algorithms.
\newblock In Gottlob et~al. \cite{DBLP:conf/gcai/2015}, pages 263--274.

\bibitem{Sch02-AICOMM}
Stephan Schulz.
\newblock {E - A Brainiac Theorem Prover}.
\newblock {\em AI Commun.}, 15(2-3):111--126, 2002.

\bibitem{Schulz13}
Stephan Schulz.
\newblock System description: {E} 1.8.
\newblock In Kenneth~L. McMillan, Aart Middeldorp, and Andrei Voronkov,
  editors, {\em LPAR}, volume 8312 of {\em LNCS}, pages 735--743. Springer,
  2013.

\bibitem{Urban06}
Josef Urban.
\newblock {MPTP} 0.2: Design, implementation, and initial experiments.
\newblock {\em J. Autom. Reasoning}, 37(1-2):21--43, 2006.

\bibitem{blistr}
Josef Urban.
\newblock {BliStr: The Blind Strategymaker}.
\newblock In Gottlob et~al. \cite{DBLP:conf/gcai/2015}, pages 312--319.

\bibitem{Veroff96}
Robert Veroff.
\newblock Using hints to increase the effectiveness of an automated reasoning
  program: Case studies.
\newblock {\em J. Autom. Reasoning}, 16(3):223--239, 1996.

\bibitem{2020SciPy-NMeth}
Pauli Virtanen, Ralf Gommers, Travis~E. Oliphant, Matt Haberland, Tyler Reddy,
  David Cournapeau, Evgeni Burovski, Pearu Peterson, Warren Weckesser, Jonathan
  Bright, St{\'e}fan~J. {van der Walt}, Matthew Brett, Joshua Wilson, K.~Jarrod
  Millman, Nikolay Mayorov, Andrew R.~J. Nelson, Eric Jones, Robert Kern, Eric
  Larson, C~J Carey, {\.I}lhan Polat, Yu~Feng, Eric~W. Moore, Jake
  {VanderPlas}, Denis Laxalde, Josef Perktold, Robert Cimrman, Ian Henriksen,
  E.~A. Quintero, Charles~R. Harris, Anne~M. Archibald, Ant{\^o}nio~H. Ribeiro,
  Fabian Pedregosa, Paul {van Mulbregt}, and {SciPy 1.0 Contributors}.
\newblock {{SciPy} 1.0: Fundamental Algorithms for Scientific Computing in
  Python}.
\newblock {\em Nature Methods}, 17:261--272, 2020.

\end{thebibliography}
\newpage

\appendix
\section{Additional Data From the Experiments}
\label{app:data}
\subsection{Parameters of the GNNs Used In Leapfrogging}

The main parameters of the graph neural networks are the number of graph convolutional
\emph{layers} used and the number of \emph{epochs} for which they were
trained. When used in ENIGMA, two additional important parameters are the number
of \emph{context} clauses used by ENIGMA-GNN and the number of
\emph{query} clauses used by ENIGMA-GNN. Table~\ref{gnntab} shows
these parameters for the leapfrogging experiments.

\begin{table}[htbp]
\begin{center}
\begin{small}
  \caption{Parameters of the GNNs used in the leapfrogging experiments}\label{gnntab}
  \centering
  \begin{tabular}{lcccc}
    \toprule
    Method & layers & epoch & context &query \\
    \midrule
    $G_1$ & 10 & 69 & 256 & 128 \\
    $G_2$ & 10 & 55 &  768 & 512 \\
    $G_3$ & 8 & 389 & 768 & 256 \\
 $G_4$ & 10 & 15 & 512 & 256 \\
    \bottomrule
  \end{tabular}
\end{small}
\end{center}
\end{table}

\subsection{Interesting Frequent Computational Patterns}

Here we show several of the computationally looking proofs found only by the Split and Merge algorithm.

\begin{figure}[htbp]
\begin{center}
\includegraphics[width=11cm]{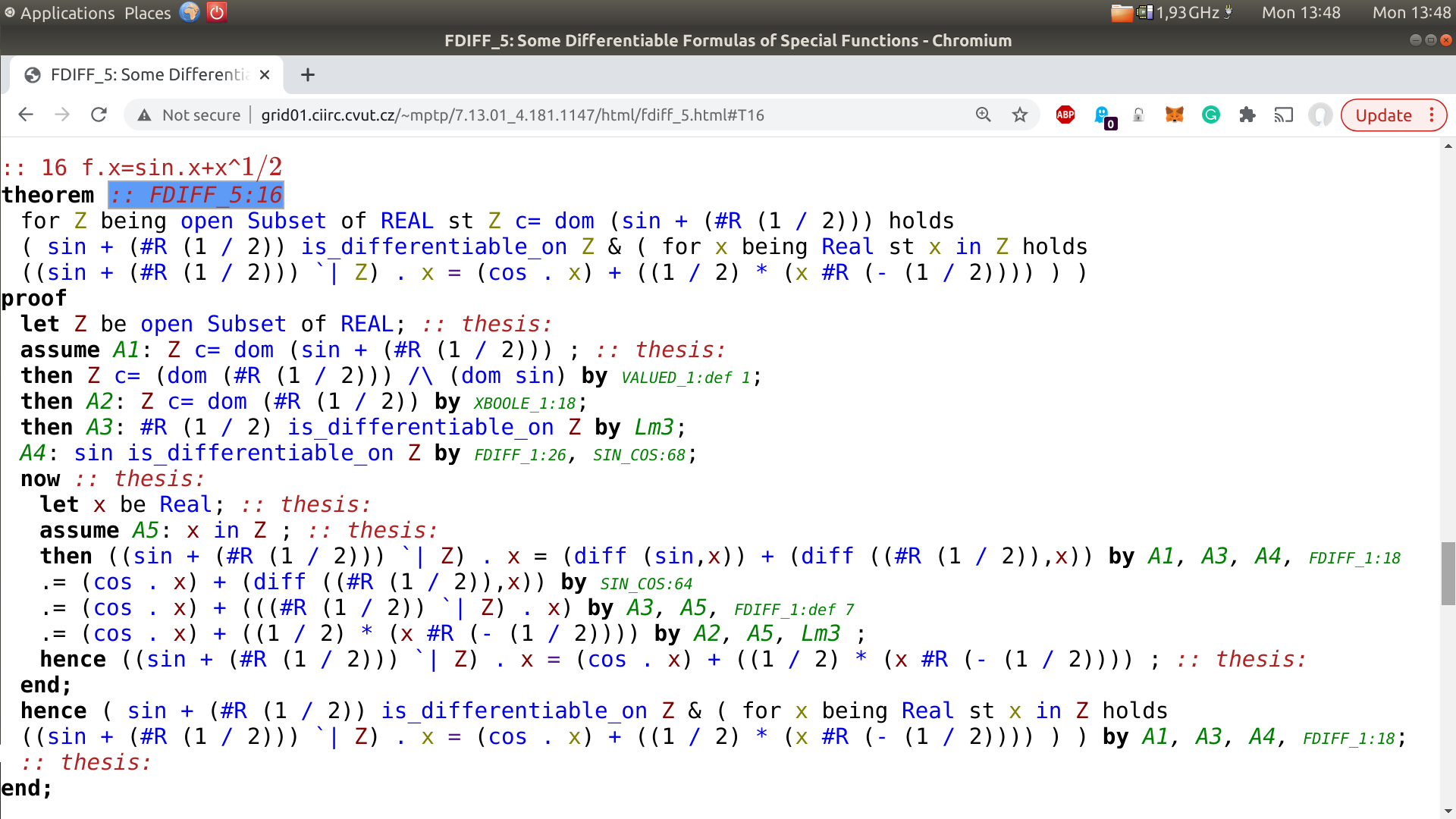}
\end{center}
\caption{Differentiation -- T16\_FDIFF\_5}\label{Differentiation}
\end{figure}

\begin{figure}[htbp]
\begin{center}
\includegraphics[width=11cm]{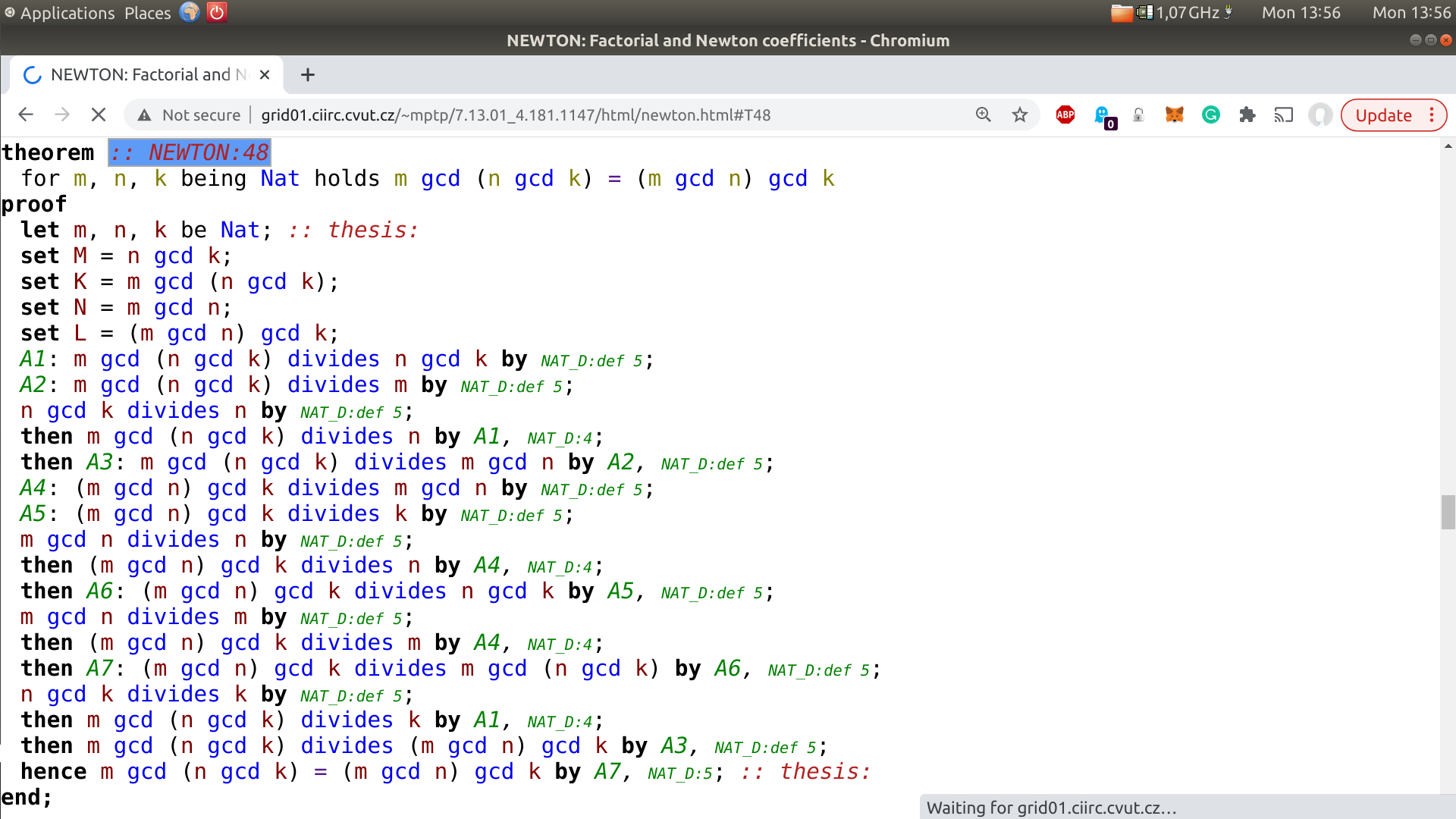}
\end{center}
\caption{Associativity of gcd by many rewrites -- T48\_NEWTON}\label{Newton}
\end{figure}

\begin{figure}[htbp]
\begin{center}
\includegraphics[width=11cm]{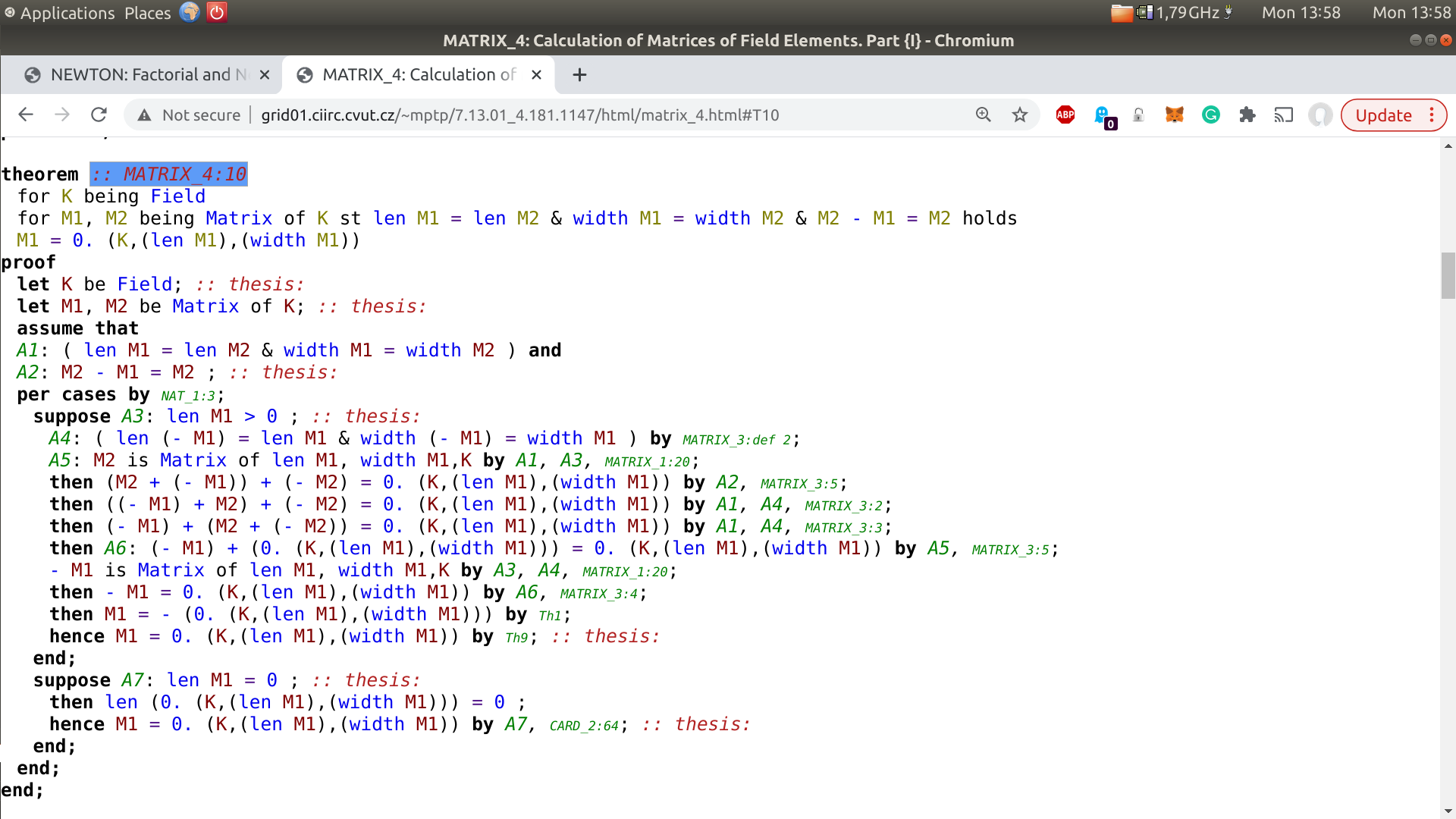}
\end{center}
\caption{Matrix manipulation -- T10\_MATRIX\_4}\label{Newton}
\end{figure}

\end{document}